\documentclass[twocolumn]{aastex7}

\shorttitle{exoALMA channel maps}
\shortauthors{Pinte et al.}
\graphicspath{{./}{figures/}}

\begin{document}

\title{exoALMA. X. channel maps reveal complex $^{12}$CO abundance distributions and a variety of kinematic structures with evidence for embedded planets}

\author[0000-0001-5907-5179]{Christophe Pinte}
\email[show]{christophe.pinte@univ-grenoble-alpes.fr}
\affiliation{Univ. Grenoble Alpes, CNRS, IPAG, 38000 Grenoble, France}
\affiliation{School of Physics and Astronomy, Monash University, Clayton VIC 3800, Australia}

\author[0000-0003-1008-1142]{John~D.~Ilee}
\affiliation{School of Physics and Astronomy, University of Leeds, Leeds, UK, LS2 9JT}
\email{fakeemail@google.com}

\author[0000-0001-6947-6072]{Jane Huang}
\affiliation{Department of Astronomy, Columbia University, 538 W. 120th Street, Pupin Hall, New York, NY, USA}
\email{fakeemail@google.com}

\author[0000-0002-7695-7605]{Myriam Benisty}
\affiliation{Universit\'e C\^ote d'Azur, Observatoire de la C\^ote d'Azur, CNRS, Laboratoire Lagrange, France}
\affiliation{Max-Planck Institute for Astronomy (MPIA), Königstuhl 17, 69117 Heidelberg, Germany}
\email{fakeemail@google.com}

\author[0000-0003-4689-2684]{Stefano Facchini}
\affiliation{Dipartimento di Fisica, Universit\`a degli Studi di Milano, Via Celoria 16, 20133 Milano, Italy}
\email{fakeemail@google.com}

\author[0000-0003-1117-9213]{Misato Fukagawa}
\affiliation{National Astronomical Observatory of Japan, Osawa 2-21-1, Mitaka, Tokyo 181-8588, Japan}
\email{fakeemail@google.com}

\author[0000-0003-1534-5186]{Richard Teague}
\affiliation{Department of Earth, Atmospheric, and Planetary Sciences, Massachusetts Institute of Technology, Cambridge, MA 02139, USA}
\email{fakeemail@google.com}


\author[0000-0001-7258-770X]{Jaehan Bae}
\affiliation{Department of Astronomy, University of Florida, Gainesville, FL 32611, USA}
\email{fakeemail@google.com}

\author[0000-0001-6378-7873]{Marcelo Barraza-Alfaro}
\affiliation{Department of Earth, Atmospheric, and Planetary Sciences, Massachusetts Institute of Technology, Cambridge, MA 02139, USA}
\email{fakeemail@google.com}

\author[0000-0002-2700-9676]{Gianni Cataldi}
\affiliation{National Astronomical Observatory of Japan, 2-21-1 Osawa, Mitaka, Tokyo 181-8588, Japan}
\email{fakeemail@google.com}

\author[0000-0003-3713-8073]{Nicolás Cuello}
\affiliation{Univ. Grenoble Alpes, CNRS, IPAG, 38000 Grenoble, France}
\email{fakeemail@google.com}

\author[0000-0003-2045-2154]{Pietro Curone}
\affiliation{Dipartimento di Fisica, Universit\`a degli Studi di Milano, Via Celoria 16, 20133 Milano, Italy}
\affiliation{Departamento de Astronom\'ia, Universidad de Chile, Camino El Observatorio 1515, Las Condes, Santiago, Chile}
\email{fakeemail@google.com}

\author[0000-0002-1483-8811]{Ian Czekala}
\affiliation{School of Physics \& Astronomy, University of St. Andrews, North Haugh, St. Andrews KY16 9SS, UK}
\email{fakeemail@google.com}

\author[0000-0003-4679-4072]{Daniele Fasano}
\affiliation{Universit\'e C\^ote d'Azur, Observatoire de la C\^ote d'Azur, CNRS, Laboratoire Lagrange, France}
\email{fakeemail@google.com}

\author[0000-0002-9298-3029]{Mario Flock}
\affiliation{Max-Planck Institute for Astronomy (MPIA), Königstuhl 17, 69117 Heidelberg, Germany}
\email{fakeemail@google.com}

\author[0000-0002-5503-5476]{Maria Galloway-Sprietsma}
\affiliation{Department of Astronomy, University of Florida, Gainesville, FL 32611, USA}
\email{fakeemail@google.com}

\author[0000-0002-5910-4598]{Himanshi Garg}
\affiliation{School of Physics and Astronomy, Monash University, Clayton VIC 3800, Australia}
\email{fakeemail@google.com}

\author[0000-0002-8138-0425]{Cassandra Hall}
\affiliation{Department of Physics and Astronomy, The University of Georgia, Athens, GA 30602, USA}
\affiliation{Center for Simulational Physics, The University of Georgia, Athens, GA 30602, USA}
\affiliation{Institute for Artificial Intelligence, The University of Georgia, Athens, GA, 30602, USA}
\email{fakeemail@google.com}

\author[0000-0003-1502-4315]{Iain Hammond}
\affiliation{School of Physics and Astronomy, Monash University, VIC 3800, Australia}
\email{fakeemail@google.com}

\author[0009-0003-7403-9207]{Caitlyn Hardiman}
\affiliation{School of Physics and Astronomy, Monash University, VIC 3800, Australia}
\email{fakeemail@google.com}

\author[0000-0001-7641-5235]{Thomas Hilder}
\affiliation{School of Physics and Astronomy, Monash University, VIC 3800, Australia}
\email{fakeemail@google.com}

\author[0000-0001-8446-3026]{Andr\'es F. Izquierdo}
\affiliation{Department of Astronomy, University of Florida, Gainesville, FL 32611, USA}
\affiliation{Leiden Observatory, Leiden University, P.O. Box 9513, NL-2300 RA Leiden, The Netherlands}
\affiliation{European Southern Observatory, Karl-Schwarzschild-Str. 2, D-85748 Garching bei M\"unchen, Germany}
\affiliation{NASA Hubble Fellowship Program Sagan Fellow}
\email{fakeemail@google.com}

\author[0000-0001-7235-2417]{Kazuhiro Kanagawa}
\affiliation{College of Science, Ibaraki University, 2-1-1 Bunkyo, Mito, Ibaraki 310-8512, Japan}
\email{fakeemail@google.com}

\author[0000-0002-8896-9435]{Geoffroy Lesur}
\affiliation{Univ. Grenoble Alpes, CNRS, IPAG, 38000 Grenoble, France}
\email{fakeemail@google.com}

\author[0000-0002-2357-7692]{Giuseppe Lodato}
\affiliation{Dipartimento di Fisica, Universit\`a degli Studi di Milano, Via Celoria 16, 20133 Milano, Italy}
\email{fakeemail@google.com}

\author[0000-0003-4663-0318]{Cristiano Longarini}
\affiliation{Institute of Astronomy, University of Cambridge, Madingley Road, CB3 0HA, Cambridge, UK}
\affiliation{Dipartimento di Fisica, Universit\`a degli Studi di Milano, Via Celoria 16, 20133 Milano, Italy}
\email{fakeemail@google.com}

\author[0000-0002-8932-1219]{Ryan A. Loomis}
\affiliation{National Radio Astronomy Observatory, 520 Edgemont Rd., Charlottesville, VA 22903, USA}
\email{fakeemail@google.com}

\author[0000-0002-9626-2210]{Fr\'ed\'eric Masset}
\affiliation{Instituto de Ciencias F\'isicas, Universidad Nacional Aut\'onoma de M\'exico, Av. Universidad s/n, 62210 Cuernavaca, Mor., Mexico}
\email{fakeemail@google.com}

\author[0000-0002-1637-7393]{Francois Menard}
\affiliation{Univ. Grenoble Alpes, CNRS, IPAG, 38000 Grenoble, France}
\email{fakeemail@google.com}

\author[0000-0003-4039-8933]{Ryuta Orihara}
\affiliation{College of Science, Ibaraki University, 2-1-1 Bunkyo, Mito, Ibaraki 310-8512, Japan}
\email{fakeemail@google.com}

\author[0000-0002-4716-4235]{Daniel J. Price}
\affiliation{School of Physics and Astronomy, Monash University, Clayton VIC 3800, Australia}
\email{fakeemail@google.com}

\author[0000-0003-4853-5736]{Giovanni Rosotti}
\affiliation{Dipartimento di Fisica, Universit\`a degli Studi di Milano, Via Celoria 16, 20133 Milano, Italy}
\email{fakeemail@google.com}

\author[0000-0002-0491-143X]{Jochen Stadler}
\affiliation{Universit\'e C\^ote d'Azur, Observatoire de la C\^ote d'Azur, CNRS, Laboratoire Lagrange, France}
\email{fakeemail@google.com}

\author[0000-0003-1412-893X]{Hsi-Wei Yen}
\affiliation{Academia Sinica Institute of Astronomy \& Astrophysics, 11F of Astronomy-Mathematics Building, AS/NTU, No.1, Sec. 4, Roosevelt Rd, Taipei 10617, Taiwan}
\email{fakeemail@google.com}

\author[0000-0002-3468-9577]{Gaylor Wafflard-Fernandez}
\affiliation{Univ. Grenoble Alpes, CNRS, IPAG, 38000 Grenoble, France}
\email{fakeemail@google.com}

\author[0000-0003-1526-7587]{David J. Wilner}
\affiliation{Center for Astrophysics | Harvard \& Smithsonian, Cambridge, MA 02138, USA}
\email{fakeemail@google.com}

\author[0000-0002-7501-9801]{Andrew J. Winter}
\affiliation{Universit\'e C\^ote d'Azur, Observatoire de la C\^ote d'Azur, CNRS, Laboratoire Lagrange, France}
\affiliation{Max-Planck Institute for Astronomy (MPIA), Königstuhl 17, 69117 Heidelberg, Germany}
\email{fakeemail@google.com}

\author[0000-0002-7212-2416]{Lisa W\"olfer}
\affiliation{Department of Earth, Atmospheric, and Planetary Sciences, Massachusetts Institute of Technology, Cambridge, MA 02139, USA}
\email{fakeemail@google.com}

\author[0000-0001-8002-8473]{Tomohiro C. Yoshida}
\affiliation{National Astronomical Observatory of Japan, 2-21-1 Osawa, Mitaka, Tokyo 181-8588, Japan}
\affiliation{Department of Astronomical Science, The Graduate University for Advanced Studies, SOKENDAI, 2-21-1 Osawa, Mitaka, Tokyo 181-8588, Japan}
\email{fakeemail@google.com}

\author[0000-0001-9319-1296]{Brianna Zawadzki}
\affiliation{Department of Astronomy, Van Vleck Observatory, Wesleyan University, 96 Foss Hill Drive, Middletown, CT 06459, USA}
\affiliation{Department of Astronomy \& Astrophysics, 525 Davey Laboratory, The Pennsylvania State University, University Park, PA 16802, USA}
\email{fakeemail@google.com}

\begin{abstract}
  We analyze the $^{12}$CO $J=3-2$ data cubes of the disks in the exoALMA program.  13/15 disks reveal a variety of kinematic substructures in individual channels: large-scale arcs or spiral arms, localized velocity kinks, and/or multiple faints arcs that appear like filamentary structures on the disk surface.  We find kinematic signatures that are consistent with planet wakes in six disks: AA~Tau, SY~Cha, J1842, J1615, LkCa~15 and HD~143006. Comparison with hydrodynamical and radiative transfer simulations suggests planets with orbital radii between 80 and 310\,au and masses between 1 and 5 M$_\mathrm{Jup}$. Additional kinematic substructures limit our ability to place tight constraints on the planet masses. When the inclination is favorable to separate the upper and lower surfaces (near 45$^\mathrm{o}$, \emph{i.e.} in 7/15 disks), we always detect the vertical CO snowline and find that the $^{12}$CO freeze-out is partial in the disk midplane, with a depletion factor of $\approx 10^{-3}$ -- $10^{-2}$ compared to the warm molecular layer. In these same seven disks, we also systematically detect evidence of CO desorption in the outer regions.

\end{abstract}

\keywords{protoplanetary disks  ---
  planet-disk interaction --- submillimeter: planetary systems --- hydrodynamics --- radiative transfer -- planets}

\section{Introduction} \label{sec:intro}
The remarkable sensitivity of the exoALMA Large Program \citep{Teague_exoALMA} enables us to map molecular line emission from protoplanetary disks with unprecedented detail. With a velocity resolution of $\approx 26\mathrm{m\,s}^{-1}$ and spatial resolution of $\approx 0\farcs1$, we can explore the intricate distribution of gas and its dynamics, which are critical for understanding the physical processes at play during planet formation  \citep{PintePPVII}. Previous papers in this series primarily focused on integrated (moment and/or peak) maps and azimuthally averaged velocity curves to search for non-Keplerian motions,
pressure gradients, or non-thermal broadening \citep{Izquierdo_exoALMA,Hilder_exoALMA,Stadler_exoALMA,Yoshida_exoALMA}.

In this paper, we extend the analysis to individual channel maps, which provide a complementary approach by isolating specific velocity channels across the emission line profiles. This allows us to search for subtle kinematic structures that could be overlooked when examining spectrally concatenated data, to resolve both the upper and lower molecular emission surfaces, and to search for diffuse and faint emission.
Such emission may trace complex chemical processes, such as the freeze-out of volatile molecules onto dust grains, photodissociation and photodesorption driven by ultraviolet light, offering insight into the disk's thermal structure and chemistry. These processes can alter the chemical makeup of the disk, influencing the composition of material that eventually forms planets \citep{Oberg2023}.

We classify the 15 exoALMA disks based on their kinematic substructures, identifying distinct patterns that suggest various underlying physical processes. For disks exhibiting signatures consistent with the presence of embedded planets, we perform hydrodynamical and radiative transfer simulations,
with the goal of estimating the mass of the potential planets.

\begin{table*}
  \caption{Main kinematic substructures in the exoALMA sources. The spatial and spectral resolutions with their corresponding peak brightness temperatures and sensitivity are indicated for the cubes used in figures \ref{fig:arcs} to \ref{fig:cont}.\label{table:obs} }
  \centering
  \begin{tabular}{l c c c c c c}
    \hline
    \hline
    Source & v$_\mathrm{syst}$ [km/s] & Beam ["] & $\Delta$v [km/s] & Peak T$_\mathrm{b}$ [K]  & T$_\mathrm{b}$ Sensitivity [K] \tablenotemark{a} & Kinematic Deviations\\
    \hline
MWC 758    & 5.86 & 0.12 &  100 & 95 & 4.5 & Large-scale arcs\\
HD 135344B & 7.07 & 0.11 &  100 & 114 & 3.4& Large-scale arcs + velocity kink\\
CQ Tau     & 6.20 & 0.12 &  200 & 91 & 1.7 &Large-scale arcs + velocity kink\\
J1604      & 4.60 & 0.15 &  100 & 84 & 1.5 & Large-scale arcs\\
AA Tau     & 6.50  & 0.15 &  100 & 73 & 1.7 & Velocity kink\\
SY Cha     & 4.10 & 0.15 &  100 & 56 & 1.6 & Velocity kink + filaments\\
J1842      & 5.93 & 0.15 &  100 & 64 & 1.5 & Velocity kink \\
J1615      & 4.75 & 0.15 &  100 & 56 & 1.3 & Velocity kink + filaments\\
LkCa 15    & 6.29 & 0.15 &  100 & 57 & 1.2 & Velocity kink + filaments\\
HD 143006  & 7.70 & 0.10 &  100 & 78 & 4.9 & Velocity kink \\
J1852      & 5.46 & 0.12 &  100 & 82 & 2.7 & Filaments \\
DM Tau     & 6.03 & 0.20 &  100 & 56 & 1.1 & Filaments + velocity kink \\
HD 34282   & -2.30 & 0.11 &  100 & 93 & 2.2 & Filaments\\
V4046 Sgr  & 2.90 & 0.12 &  100 & 98 & 2.3 & Smooth\\
PDS 66     & 3.96 & 0.15 &  100 & 81 & 1.8 & Smooth\\
    \hline
  \end{tabular}
  \tablenotetext{a}{We used the Rayleigh-Jeans approximation for sensitivity to maintain consistency with the ALMA observing tool, while the full blackbody function was used for all other brightness temperatures.}
\end{table*}

\begin{figure*}
  \includegraphics[width=\linewidth]{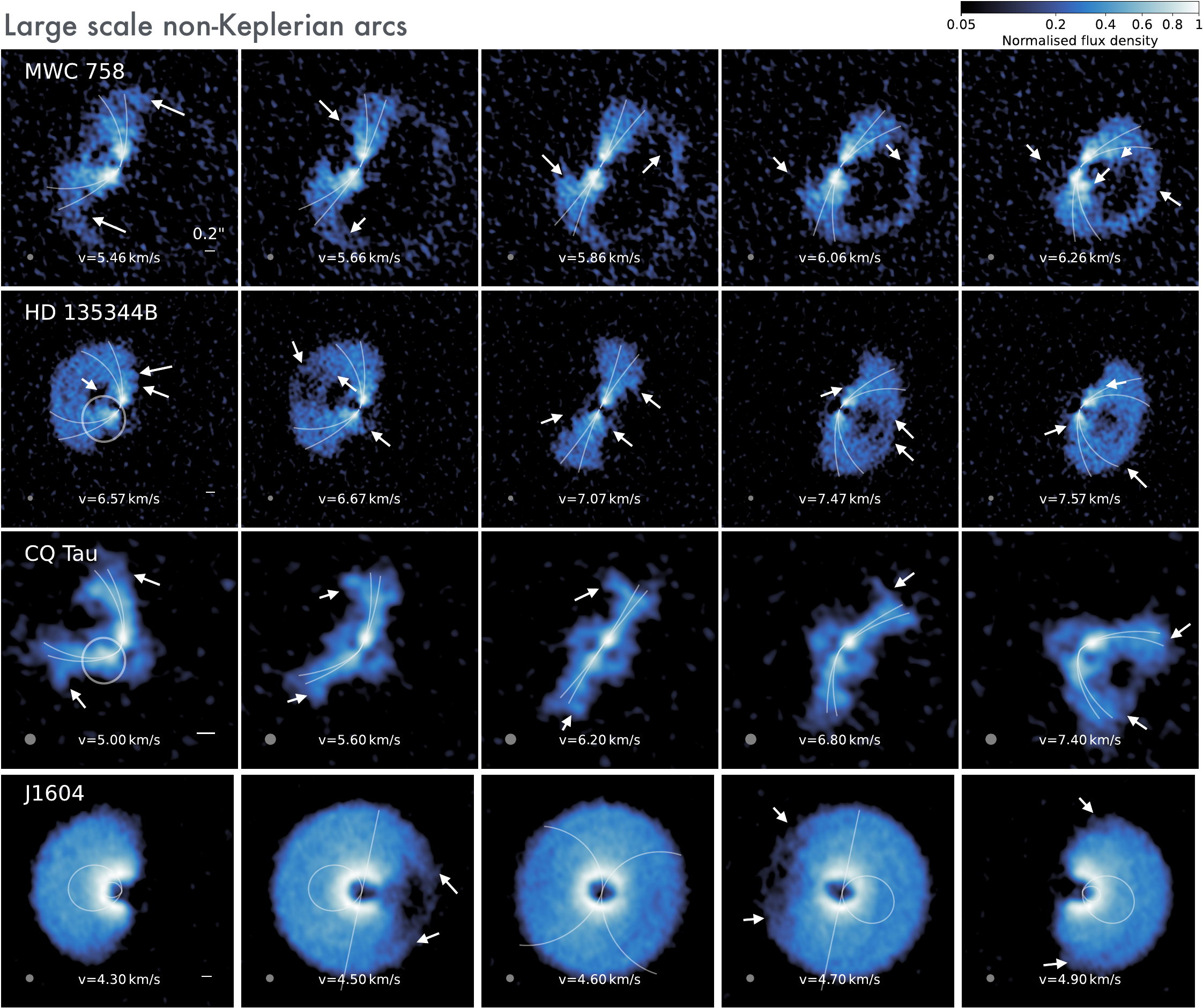}
  \caption{Sources displaying large-scale non-Keplerian arcs. For each object, we show the channel at systemic velocity (central column), as well as channels at two velocity offsets on the blue side of the line (first two columns) and the opposite channels on the red side of the line (last two columns).  Solid lines indicate the expected location of the isovelocity curves at Keplerian velocities $\pm$100\,m\,s. White arrows indicate deviations from the expected isovelocity curves for Keplerian rotation, and white circles highlight velocity kinks, \emph{i.e.} distortions of the isovelocity curves. The beam is indicated as a gray circle and the horizontal white line is a $0\farcs2$ scale. The color map traces the square root of the intensity between zero and the peak. Note that the ringlike structure seen in CQ~Tau is the continuum emission. \label{fig:arcs}}
\end{figure*}

\begin{figure*}
  \includegraphics[width=\linewidth]{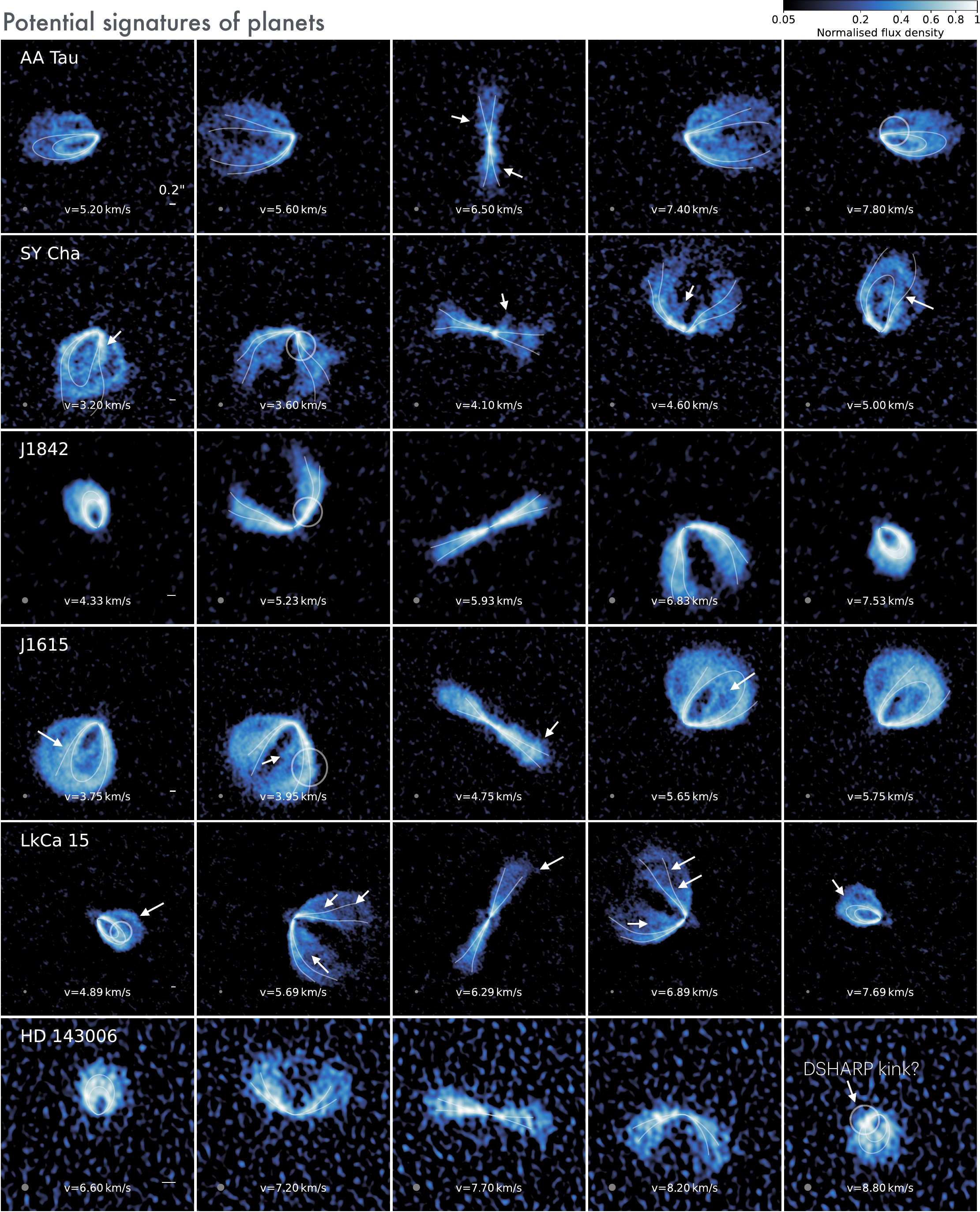}
  \caption{Sources with velocity kinks consistent with embedded planets (white circles). White arrows indicate additional deviations from Keplerian rotation. Beams and scales as in Fig~\ref{fig:arcs}. \label{fig:kinks}}
\end{figure*}

\begin{figure*}
  \includegraphics[width=\linewidth]{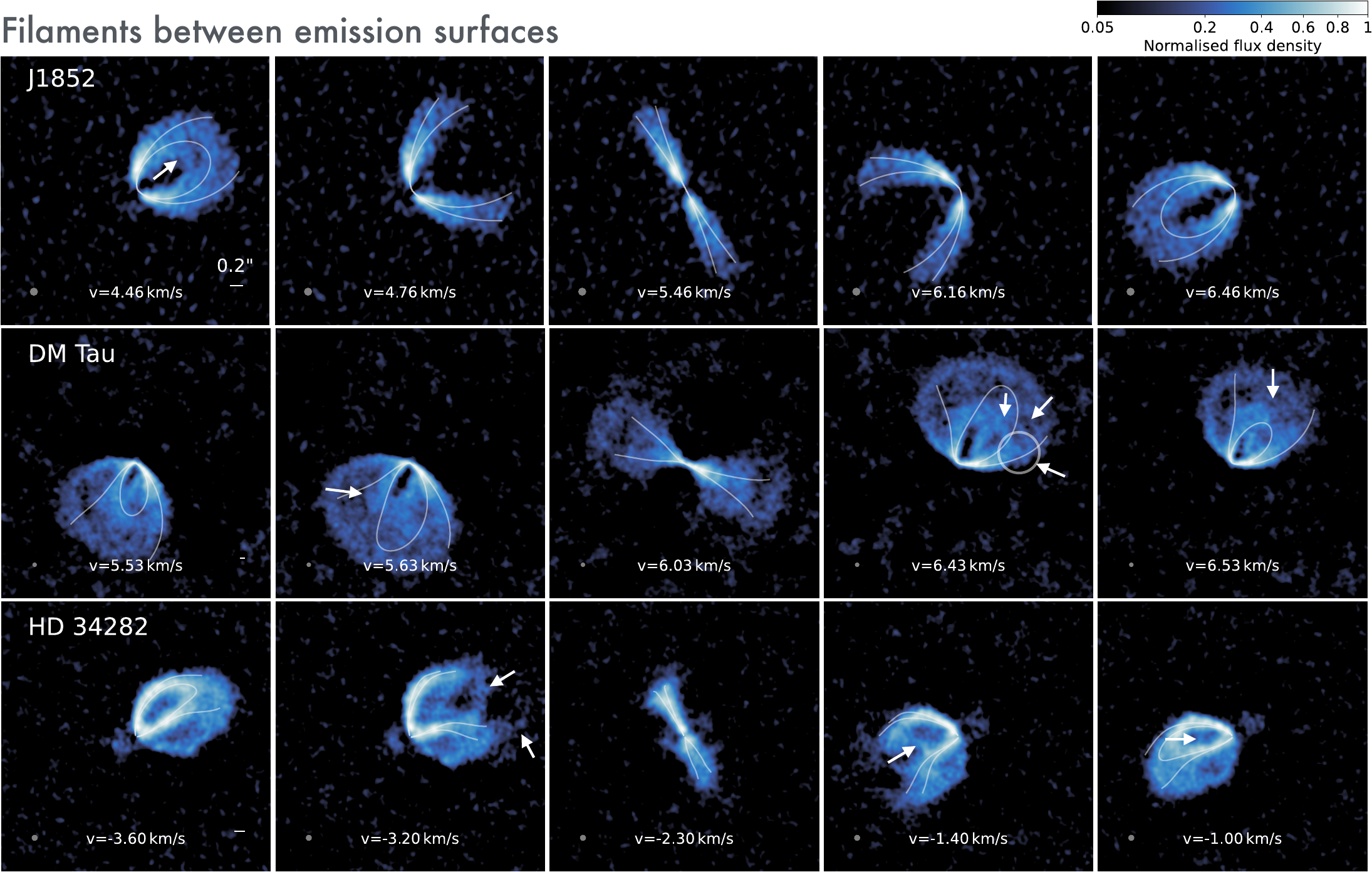}
  \caption{Sources showing a filamentary pattern between the emission surfaces (indicated by white arrows). Note that such patterns are also visible in SY~Cha, J1615 and LkCa~15, as shown in Fig.~\ref{fig:kinks}. A potential kink in the isovelocity is shown with a white circle.
    Beams and scales as in Fig~\ref{fig:arcs}. \label{fig:filaments}}
\end{figure*}

\begin{figure*}
  \includegraphics[width=\linewidth]{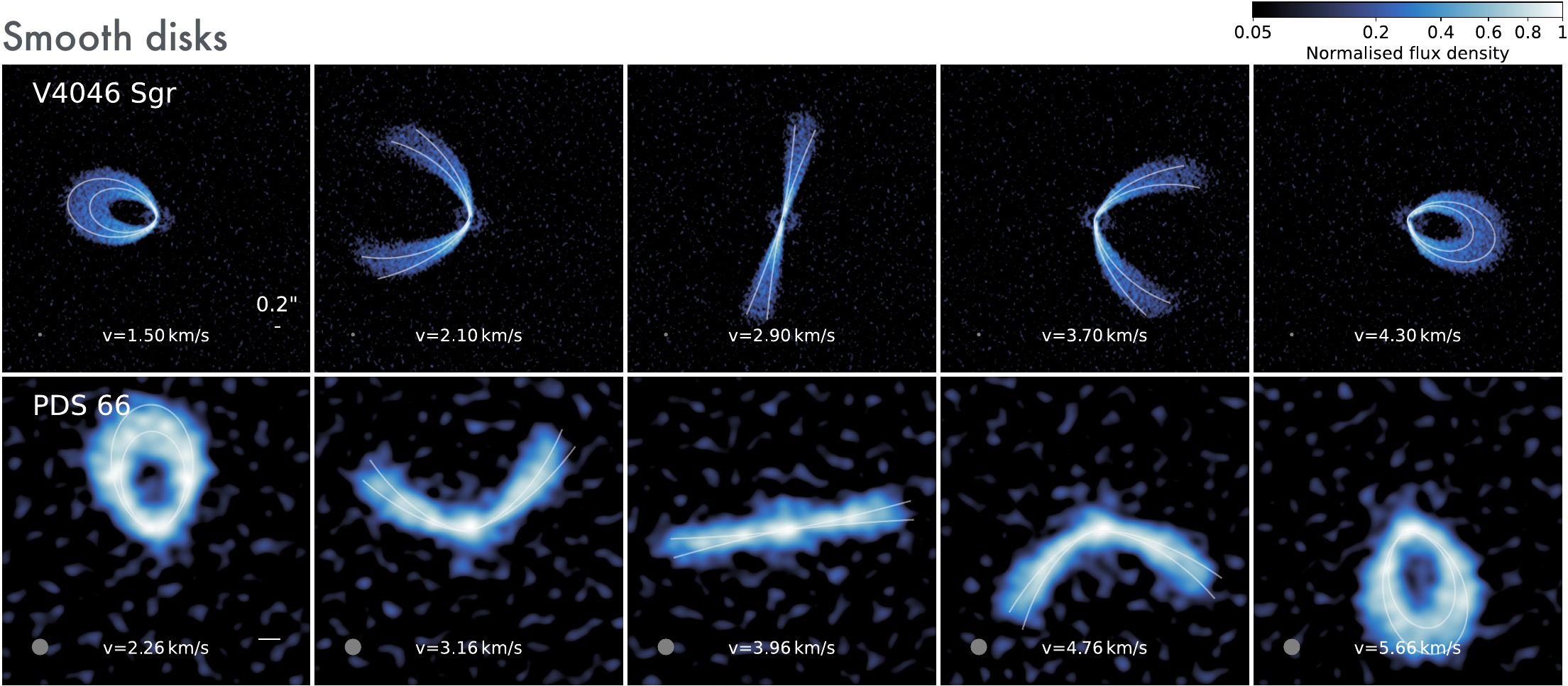}
  \caption{Sources with smooth, Keplerian-looking, velocity structure, or with signal-to-noise too low to extract reliable structures at the considered scales. Beams and scales as in Fig~\ref{fig:arcs}. \label{fig:smooth}}
\end{figure*}

\section{Observations and Imaging}

We inspected the $^{12}$CO $J=3-2$ exoALMA data cubes at various spatial ($0\farcs1$ to $0\farcs3$) and spectral (28 to 200\,m/s) resolutions to systematically search for kinematic substructures. In most cases, we also reimaged the visibilities using the standard exoALMA imaging procedure \citep{Loomis_exoALMA},  but with additional weighting schemes and spectral binning, in order to tailor the spatial resolution and signal-to-noise ratio, and ascertain the robustness of the detected kinematic structures (Table~\ref{table:obs}).
Continuum subtraction may impact the measured brightness temperature and apparent morphology of optically thick line emission \citep{Boehler2017}. To ensure that the kinematic structures are not artificially created by continuum subtraction, we imaged all the channel maps without continuum subtraction, and checked that the detected kinematic deviations are also present in the corresponding continuum-subtracted maps.
The disks with the most interesting structures were also reimaged using regularized maximum likelihood imaging instead of CLEAN \citep{Zawadzki_exoALMA}, confirming that the detections presented here are not artifacts of the imaging process.

We focus in this letter on the $^{12}$CO data cubes, but also note that the $^{13}$CO cubes show counterparts of the kinematic substructures presented here, while the limited signal-to-noise ratio of CS cubes makes it difficult to robustly detect any velocity deviation.

\section{Results}

\subsection{A Variety of Kinematic Substructures}

We visually detect velocity deviations in $^{12}$CO channel maps in 13 of the 15 exoALMA objects. Figures~\ref{fig:arcs} to \ref{fig:smooth} show representative channel maps of the $^{12}$CO emission for all disks, classified according to the dominant type of kinematic structures:
(i) large-scale kinematics deviations over most of the velocity range, (ii) a primary velocity kink suggesting at least one embedded planet, (iii) multiple faints arcs that appear like a filamentary structure on the disk surface, and (iv) smooth disks with no obvious kinematic structures.
The detected kinematic substructures are summarized in Table~\ref{table:obs}.

To help visualize the velocity deviations, we overlay the expected locations of the isovelocity curves on the upper surface, assuming the disks are in Keplerian rotation. We used the code {\sf dynamite} \citep{Pinte2018a} to extract the disk’s geometric parameters (outer radius, inclination, position angle (PA), and systemic velocity, and stellar masses) and the altitude of the CO emission layer.  Parameters are  summarized in Table~\ref{table:models}.
We verified that all those parameters are consistent with the values derived with {\sf discminer} by \cite{Izquierdo_exoALMA} and that the emission surfaces closely match those derived and discussed in detail by \cite{Galloway_exoALMA}.

Figure~\ref{fig:arcs} shows sources with large-scale kinematic deviations, in the form of extended arcs or spirals, that span most of the velocity range.
Deviations are dominated by a major structure near the outer edge of the detected $^{12}$CO emission in CQ~Tau and J1604, while MWC~758 and HD~135344B also show multiple arcs at all radii, reminiscent of structures seen, for instance, in HD~142527 \citep{Garg2021} and HD~100546 \citep{Norfolk2022}. Interestingly,
CQ~Tau also displays a ``velocity kink'' located on top of the continuum ring (left panel at v=5 km/s), similar to the one detected in HD~100546 \citep{Norfolk2022}, which appears as a Doppler flip in the peak velocity residual maps \citep{Casassus2019}. A large velocity kink is also detected in HD~135344B at $\approx$ 6.6 km/s, and corrresponds to the Doppler flip discussed by \cite{Izquierdo_exoALMA}.
The large velocity deviations observed in MWC~758, HD~135344B and CQ~Tau are reminiscent of the predicted kinematic signatures from massive (above the planetary regime) companions on misaligned orbits \citep[e.g.][]{Price2018,Calcino2023,Calcino2024,Ragusa2024}. This would be consistent with the extended spiral arms seen in scattered light in these sources \citep{Muto2012,Benisty2015,Hammond2022}.
MWC~758’s outer arc shows a remarkable red-shifted emission structure with a velocity shift larger than the expected Keplerian velocity at this location, suggesting a structure not associated with the disk rotation. This may trace a wind or the remains of a past interaction with a cloud or a stellar flyby \citep{Cuello2020}. We note that \cite{Reggiani2018} found a point source at $2\farcs3$, but concluded it is likely to be a background source.
 We also note that the high-velocity channels of J1604 (not shown here) present a significant change in PA, \emph{i.e.} a warp, as previously observed by \cite{Stadler2023}, which suggested the presence of a planet inside the inner cavity (see also Fig.~6 of \citealp{Teague_exoALMA}). The almost face-on orientation of J1604 suggests that the arc seen in most channels is primarily dominated by vertical motions, which is reminiscent of buoyancy spirals \citep[e.g.][]{Bae2021}, and may indicate the presence of an embedded planet.

 Figure~\ref{fig:kinks} shows sources with velocity ``kinks'' that may indicate the wake of embedded planets. In each panel, the white circle highlights the  velocity kink near the planet candidate, but most sources also display additional deviations from Keplerian rotation, as indicated by the white arrows. The most prominent kink is detected in SY~Cha at 3.6 km/s, but additional structures are also detected at different velocities in that source. Isolating a main kink is more difficult for AA~Tau, J1842, J1615, and LkCa~15, as multiple kinks are detected across several channels. This seemingly argues against a planetary origin, as previous studies have suggested that an embedded planet would generate only localized velocity perturbations \citep{Pinte2018b,Pinte2019}. However, it was later shown that a kink is created each time the planet wake crosses an isovelocity curve \citep{Bollati2021}, as observed in HD~163296 \citep{Calcino2022} and suggested in IM~Lupi (\citealp{Verrios2022}, but we note that \citealp{Lodato2023} proposed gravitational instability (GI) as an alternative explanation for the spiral structure).
The detected velocity kinks indeed appear to be connected to large substructures seen in the peak intensity maps, as illustrated in Figure~\ref{fig:cont}. In particular, the locations of the velocity kinks align with dips in the peak brightness map for LkCa~15, SY~Cha, and J1615.
 We detect a kink in HD~143006 at 8.8km/s, at the same location and velocity as the one detected in the DSHARP data in band 6 \citep{PerezL2018,Pinte2020}. This potential kink is only visible in cubes imaged at a spatial resolution of $0\farcs1$. This is the maximum resolution attainable with our baselines, and, as a result, the limited signal-to-noise ratio makes it difficult to fully confirm the DSHARP detection. We investigate in section~\ref{sec:models} whether the observed structures might be due to planets.

\begin{table*}[!t]
  \caption{Parameters used in the {\sf phantom} + {\sf mcfost} modeling. Stellar masses and disk geometry were derived with {\sf dynamite}. The planet masses were estimated by comparing the $^{12}$CO synthetic cubes to the data (Fig.~\ref{fig:mcfost+phantom}). \label{table:models}}
  \centering
  \begin{tabular}{l c c c c c c c c}
    \hline
    \hline
    Source & M$_*$  [M$_\odot$] & M$_\mathrm{disk}$  [M$_\odot$] & i  [$^{\rm o}$] & PA  [$^{\rm o}$] & r$_\mathrm{out}$ [au] & r$_\mathrm{c}$ [au] & r$_\mathrm{planet}$ [au] & Planet Mass Estimate [M$_\mathrm{Jup}$]\\
    \hline
    AA Tau  &   0.8 & $5 \times 10^{-3}$   & 120    & 272   & 240 & 100 & 80  & 2 \\
    J1615   &  1.15 & $5 \times 10^{-3}$   & 133    & 325.5 & 435 & 250 & 310 & 2 \\
    J1842   &   1.1 & $5 \times 10^{-3}$   &  38    & 207  & 220 & 100 &  105 & 1 \\
    LkCa 15 &  1.15 &  $10^{-2}$           &  50    &  62   & 650 & 300 & 240 & 5 \\
    SY Cha  &   0.8 &   $10^{-2}$          & 52     & 348   & 310 & 150 & 140 & 5 \\
    \hline
  \end{tabular}
\end{table*}

 Figure~\ref{fig:filaments} presents sources showing filamentary structures between the main isovelocity curves, either in the form of multiple arcs in the upper emission layer (J1852 and HD~34282) between the upper and lower surfaces in the case of DM~Tau. SY Cha, J1615, and LkCa~15 (in Fig.~\ref{fig:kinks}) also show some evidence of filaments between their upper and lower emission surfaces. These structures may trace spiral arms running through the disk surface. The filamentary structure in DM Tau is detected in most channels, but is most obvious around 6.4km/s, where we also detect a tentative kink in the isovelocity curve, which appears connected to the filaments.

Finally, Figure~\ref{fig:smooth} shows sources for which we did not detect any obvious structure in the channel maps, either because the emission is smooth and appears to follow an almost Keplerian rotation profile for V4046~Sgr or because the small disk size makes it difficult to confidently detect substructures for PDS 66. The smooth velocity field, in particular compared to other exoALMA sources, is surprising as V4046~Sgr is a known spectroscopic binary \citep{Byrne1986,Stempels2004}, with known substructures in scattered light and millimiter continuum emission \citep{Avenhaus2018,Martinez-Brunnera2022}.

Our classification is subjective, and some sources clearly exhibit multiple types of substructures.  This classification is also biased by inclination, where sources near 45$^\circ$ offer the best orientation to detect small velocity perturbations that may be caused by planets or emission between the upper and lower layers, while sources closer to pole-on are more favorable for detecting vertical motions.

The structures we detect in channel maps mostly correspond to those observed in integrated or peak intensity and velocity maps as presented by \cite{Izquierdo_exoALMA}. A comparison with these integrated maps reveals that some of the kinematic deviations we see in channels are part of larger-scale kinematic structures, such as spirals, for instance in the disks of MWC~758, HD~135344B, and CQ~Tau.

\subsection{CO Vertical Snowlines and Desorption}
The channel maps also reveal a complex distribution of CO abundance within the disks, especially in those at intermediate inclinations, where projection effects between the upper and lower emitting surfaces are limited, allowing for a clearer view of the spatial distribution of the molecular emission.

First, we observe a clear separation between the upper and lower CO surfaces, highlighting the stratified structure of the emission, as has been seen in many disks already \citep{de-Gregorio-Monsalvo2013,Rosenfeld2013,Pinte2018a,Law2021,Law2022,Teague2021,Paneque-Carreno2023}. For a detailed extraction and discussion of the emission surfaces in the exoALMA sample, we refer to \cite{Galloway_exoALMA}.

Second, in every disk where we can clearly separate the upper and lower surfaces, we consistently detect non-zero emission between these surfaces. Figure~\ref{fig:CO} shows channels of HD~34282, DM~Tau, LkCa~15, J1615, SY~Cha, AA~Tau, and J1842 with a larger beam size of $0\farcs25$ to enhance fainter emission. In the central two columns, the regions between the upper and lower surfaces are indicated with orange arrows. These regions are always brighter than the areas between the near and far sides of the upper surface, indicated by red arrows. We selected channels where the projected separation between the various isovelocity curves is similar, ruling out beam effects as the cause of this difference. This confirms that the emission between the upper and lower surfaces is real, indicating that CO is not fully frozen out onto dust grains in the disk midplane, and suggesting some degree of desorption.
We note that the majority of disks exhibiting evidence of desorption also show potential signs of planets. This is likely a coincidence, resulting from favorable inclinations that allow for a clear separation of the upper and lower surfaces, in turn enabling the detection of both midplane emission and velocity kinks on the emission surfaces.

Finally, at large radii, we detect diffuse emission where the upper and lower surfaces appear to merge (white arrows in the first and last columns of Fig.~\ref{fig:CO}). This diffuse emission is also detected when extracting the emission surfaces \citep{Galloway_exoALMA}. This implies that CO remains in the gaseous phase throughout the disk's vertical extent at these distances, likely due to photo-desorption by UV radiation (either from the central star via scattering or due to external illumination). This partial CO freeze-out in the midplane and desorption in the outer regions was also observed in IM Lupi \citep{Pinte2018a}.

\section{A Planetary Origin ?}
\label{sec:models}

\subsection{Modeling}

To determine whether embedded planets explain the observed velocity kinks in Fig.~\ref{fig:kinks}, we performed a series of hydrodynamical and radiative transfer simulations for AA~Tau, LkCa~15, SY~Cha, J1615, and J1842. We note that \cite{Ballabio2021} already modeled HD~143006. Our aim is not to develop a detailed model for each source; instead, we test, through a restricted exploration of the parameter space, whether a single planet can reasonably explain the observed channel maps. In particular, we focus on the $^{12}$CO channel maps and do not attempt to model all observables, such as submillimeter continuum images or scattered light images.

We model each object as a disk surrounding a central star with one embedded planet. We used the stellar mass and geometrical parameters derived with {\sf dynamite} (Table~\ref{table:models}).
We performed 3D gas-only simulations with the Smoothed Particle Hydrodynamics (SPH) code {\sf phantom} \citep{Price2018}. We used $10^6$ particles and evolved the system for 20 planet orbits, which is sufficient to establish the flow pattern around the planet.
Because the $^{12}$CO emission arises from the disk’s upper layers, which are poorly sampled by the SPH method (where the resolution follows the mass), we split each SPH particle into 13. We place 12 new particles on an icosahedron centered on the initial particle, at a radius equal to the smoothing length of the initial particle. Each new SPH particle has a mass of 1/13 of the initial particle mass and retains the initial particle’s velocity. We evolve this higher-resolution simulation for an additional 2 planet orbits. To achieve a mean Shakura-Sunyaev viscosity of $5 \times 10^{-3}$, we set the shock viscosity to $\alpha_{\rm av} = 0.2$ which remains above the lower bound of $\alpha_{\rm av} \approx 0.1$ needed to resolve physical viscosity in {\sf phantom} \citep{Price2018}.
We set the inner radius to 5\,au.  We neglect self-gravity and use a  disk mass of $5 \times 10^{-3} \, \mathrm{M}_\odot$ for all disks, but note that we rescale the disk masses in {\sf mcfost} to the values indicated in  Table~\ref{table:models}) to match the separation between emitting layers.

 We setup the disks with a tapered surface density profile:
$$
\Sigma(r) = \Sigma_c \left( \frac{r}{r_c}\right)^{-\gamma} \exp\left( -\left( \frac{r}{r_c}\right)^{2-\gamma} \right)
$$
with $\gamma=0.8$. We adopted a vertically isothermal equation of state with $H/R=0.1$ at $r=50$\,au and a power-law index for the sound speed of -1/3.

For each object, we embedded a single planet at the projected location of the detected primary kinks with an initial mass of 1, 2, 5, or 10\,M$_\mathrm{Jup}$, and with an accretion radius set to 0.125 times the Hill radius.

\begin{figure*}
  \includegraphics[width=\linewidth]{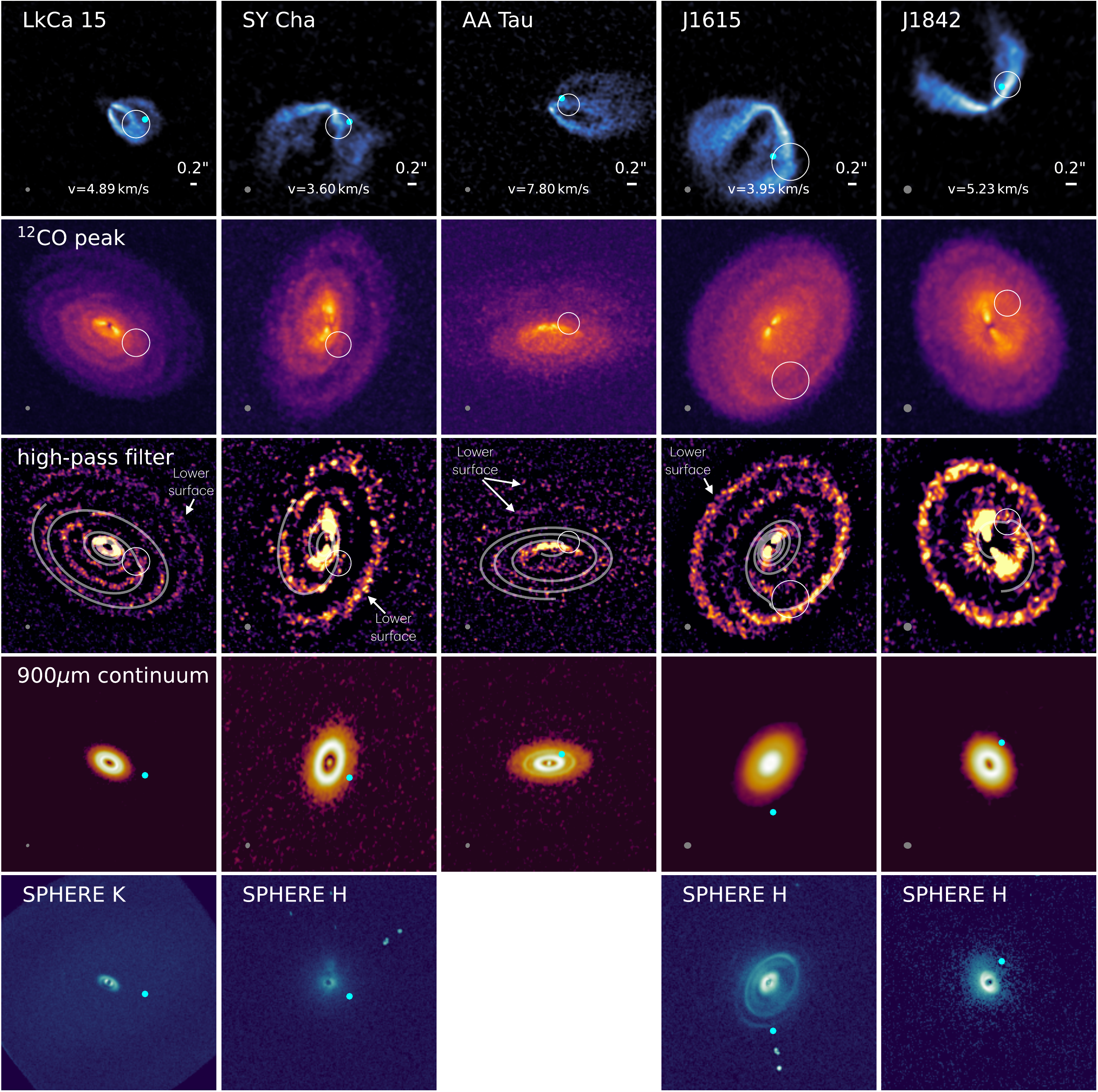}
  \caption{
$^{12}$CO $J=3-2$ observations from the Atacama Large Millimeter/submillimeter Array (ALMA), continuum emission, and near-infrared scattered light images for the five newly detected planet candidates. The first row shows the $^{12}$CO $J=3-2$ channels with the identified velocity kinks. The second row presents the corresponding peak intensity map. The third row displays a high-pass filtered peak intensity map using a Gaussian kernel twice the size of the beam, where we overlay the location of the planetary wake, derived using the same thermal structure as in the {\sf phantom} models and projected vertically onto the CO-emitting surface. The fourth row shows the exoALMA continuum \citep{Curone_exoALMA}, and the last row presents SPHERE DPI observations. The white circles indicate the velocity kinks shown in Fig.~\ref{fig:kinks}, while the blue dot marks the deprojected location on the disk midplane. No SPHERE data are available for AA~Tau, as it is currently too faint to lock the adaptive optics system.
\label{fig:cont}}
\end{figure*}

\begin{figure*}
  \centering
  \includegraphics[width=0.7\linewidth]{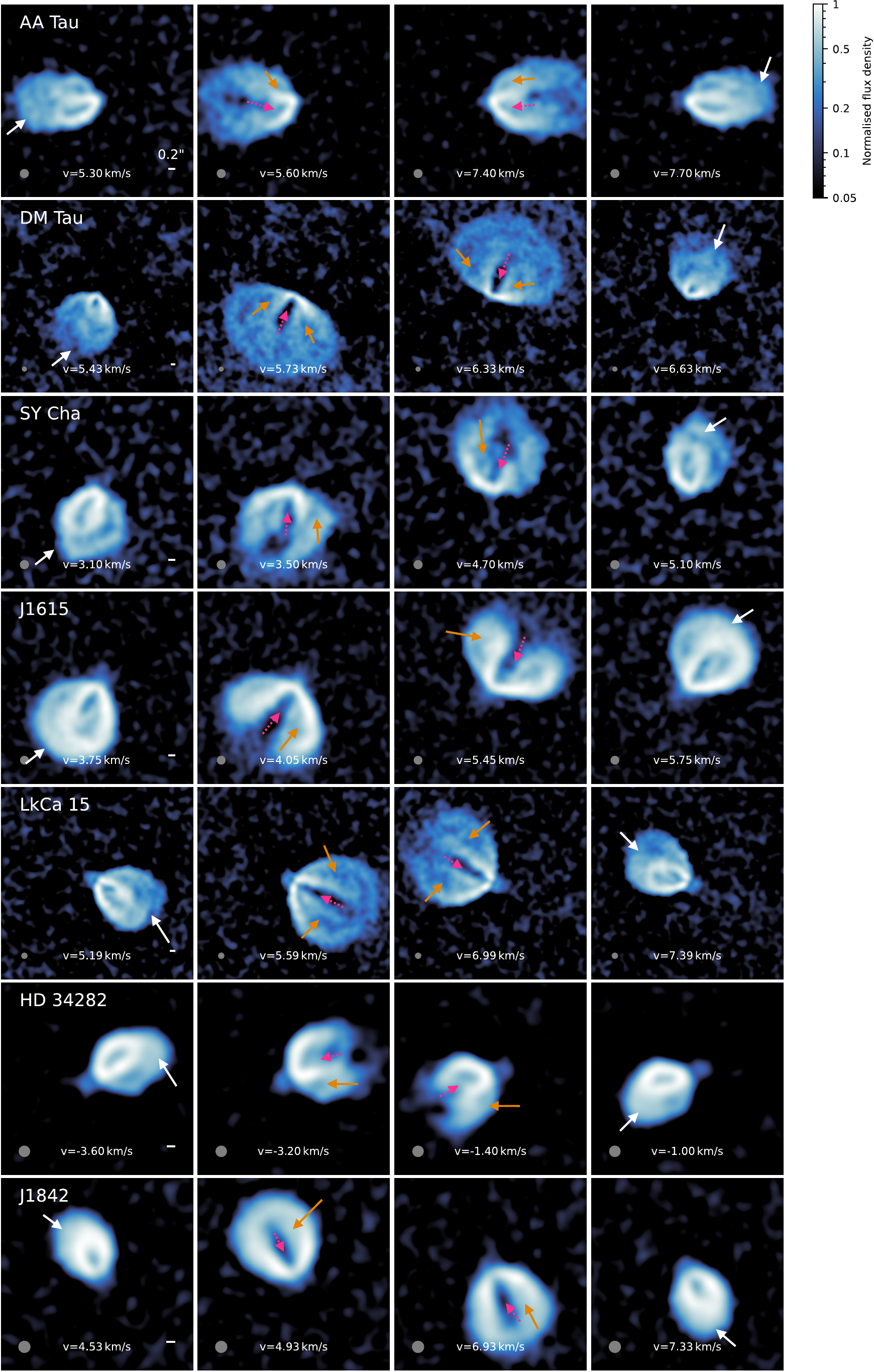}
  \caption{Evidence for complex CO abundance distribution. The white arrows in the first and last columns show regions where the upper and lower surfaces merge, indicating desorption. In the central two columns, the orange arrows show that the emission between the upper and lower surface is always higher than the emission between the near and far side of the upper isovelocity curves (red arrows with dotted line), even if the projected separation is similar. Beam is $0\farcs25$. Color map is representing the log of the intensity to highlight low-brightness regions.\label{fig:CO}}
\end{figure*}

We postprocessed the models with the radiative transfer code {\sf mcfost} \citep{Pinte2006,Pinte2009} to compute the dust temperature structure and CO maps, matching Voronoi cells to SPH particles. The combination of  {\sf phantom} and {\sf mcfost} was benchmarked and validated by \cite{Bae_exoALMA}. We assumed 1\,Myr isochrones to set the luminosities and effective temperatures of the star \citep{Siess2000} and planets \citep{Allard2012}.
As we do not see any evidence of local heating in the exoALMA data, we did not include any accretion luminosity, unlike \cite{Pinte2023}.
We assumed astrosilicate grains \citep{Weingartner2001} with sizes following
$\mathrm{d}n(a) \propto a^{-3.5}\mathrm{d}a$ between 0.03 and
1000$\mu$m, and a uniform gas-to-dust ratio of 100. We computed the dust optical properties using the Mie theory.
The dust model is only used to set the disk thermal balance, where we assumed that $T_\mathrm{gas} = T_\mathrm{dust}$, and that the CO $J=3-2$ transition is in local thermodynamic equilibrium. We set the turbulent velocity to zero, and only thermal broadening is contributing to the local line width.

We set the CO abundance following the prescription in Appendix B of \cite{Pinte2018a} to account for freeze-out, photo-dissociation, and photo-desorption. In the warm molecular layer, we set the $^{12}$CO abundance to $10^{-4}$ relative to H. Below 20\,K, we deplete this abundance by a factor that we allow to vary, unless the UV field computed by {\sf mcfost} is high enough for photo-desorption to occur.

The {\sf phantom} and {\sf mcfost} models are available in \cite{DVN/H5QUQT_2025}.

\subsection{Comparison with Data}

\begin{figure*}
  \includegraphics[width=\linewidth]{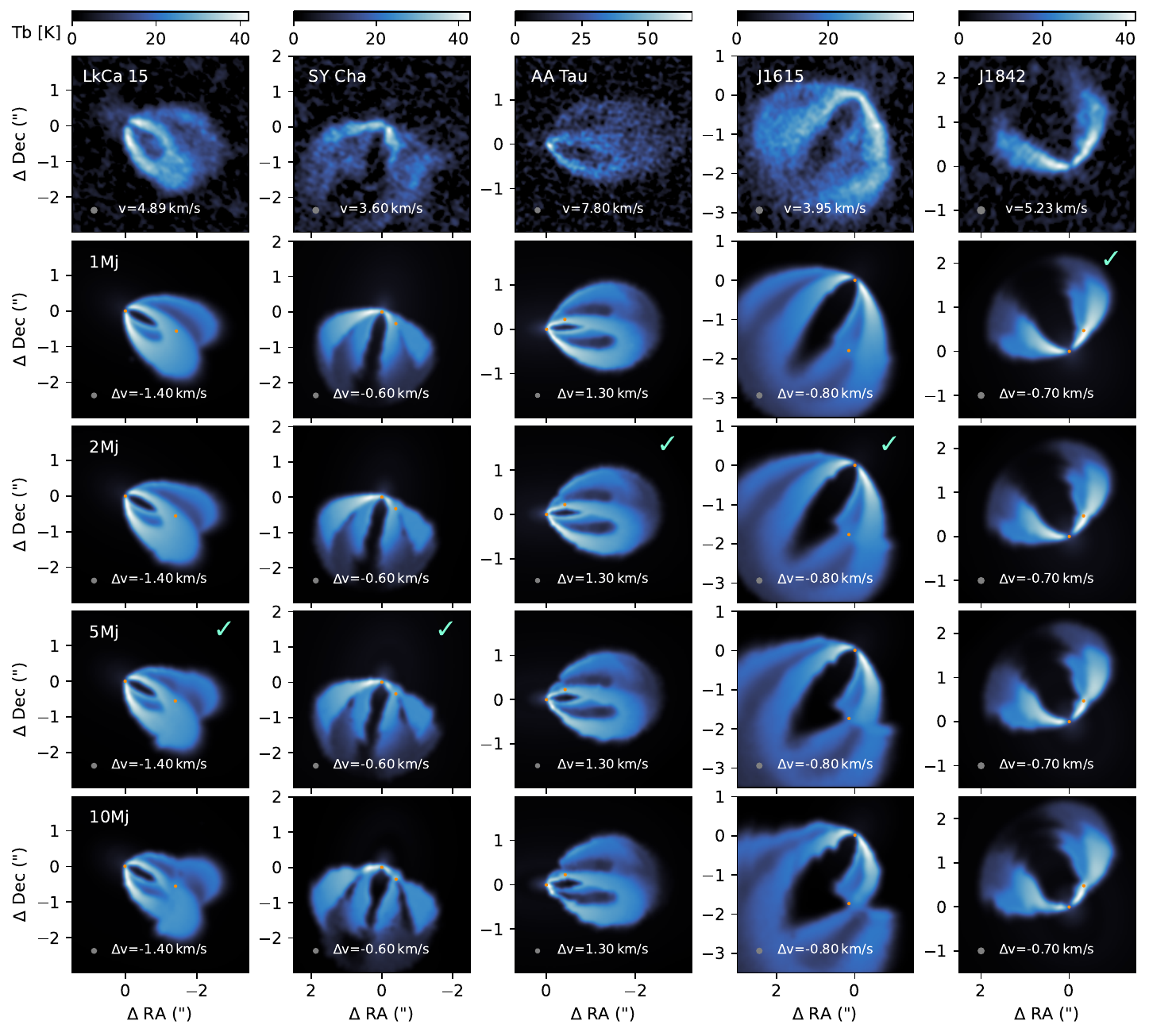}
\caption{Comparison of $^{12}$CO J=3--2 ALMA observations (top row) with synthetic channel maps
    from our 3D hydrodynamics calculations with
  embedded planets of 1, 2, 3 and 5 M$_\mathrm{Jup}$ (from top to bottom). The channel
  width is 0.1km s$^{-1}$. Synthetic maps were convolved to a Gaussian beam to
  match the spatial resolution of the observations. Orange dots show the location of the sink particles: central star and planet. The $\Delta v$ value indicates the offset from systemic velocity. The green checkmark indicates the models that visually best reproduce the observations.
\label{fig:mcfost+phantom}}
\end{figure*}

Figure~\ref{fig:mcfost+phantom} shows the predicted emission for our model grid. For each source, we compare the channel where the velocity kink is most visible to our synthetic models. In all cases, a 10\,M$_\mathrm{Jup}$ planet produces velocity deviations that are larger than those observed. The detected velocity kinks are best reproduced with a planet mass of 2\,M$_\mathrm{Jup}$ for LkCa~15, AA~Tau, and J1615, 5\,M$_\mathrm{Jup}$ for SY~Cha, and 1\,M$_\mathrm{Jup}$ for J1842.

The planet model matches the observations well for LkCa~15, J1615, and J1842, in terms of both shape and amplitude of the kink. In LkCa~15, additional planets are likely present at small separations, as discussed by \cite{Gardner_exoALMA}, who imaged the disk at higher spatial resolution ($\lesssim 0\farcs05$) by combining exoALMA data with longer-baseline observations. In particular, \cite{Gardner_exoALMA} detected an additional velocity kink at the systemic velocity at a separation of $\lesssim 0\farcs1$, which is smeared out at our spatial resolution. The agreement between our synthetic maps and the observations is particularly remarkable for J1615, where our model also reproduces the velocity kink seen on the lower surface.

For SY~Cha, a planet of approximately 5\,M$_\mathrm{Jup}$ produces a velocity kink with the correct amplitude, but its shape is not as sharp as in the observations, suggesting that other physical processes might be at play. For AA~Tau, the signal-to-noise ratio at the location of the velocity kink is low ($\approx$5), making it difficult to assess the model’s ability to precisely predict the planet mass. Nevertheless, a 10 M$_\mathrm{Jup}$ planet can be ruled out, as it would open a deep gap that causes the emission along the isovelocity curve in the southern region to appear too narrow.

\begin{figure*}
  \includegraphics[width=\linewidth]{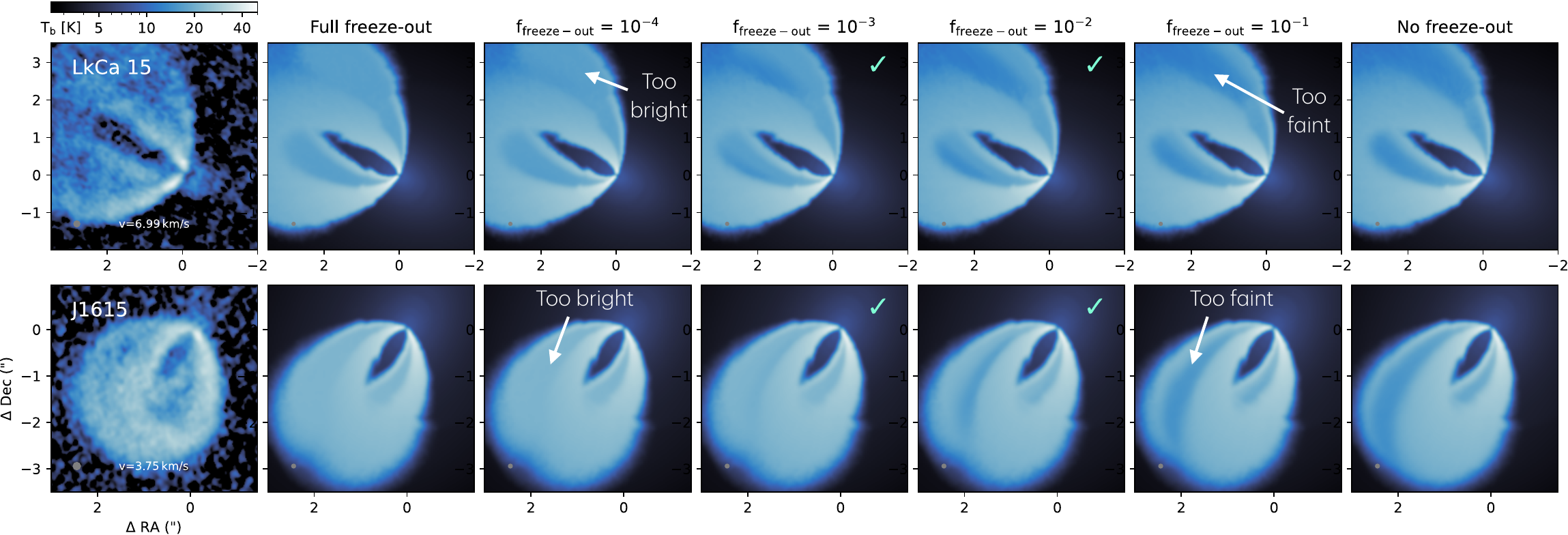}
\caption{Comparison of $^{12}$CO observations of LkCa~15 and J1615 (left column) with synthetic channel maps
  with various freeze-out depletion, from full freeze-out (\emph{i.e.}, CO abundance is 0 where the temperature is below 10\,K) to no freeze-out (\emph{i.e.}, CO abundance is the same as in the disk surface: $10^{-4}$, even where T $<20$\,K). For both sources, we used the best hydrodynamical model from Fig.~\ref{fig:mcfost+phantom}. The color map scales with the log of the brightness temperature to highlight the emission from the disk midplane where the models differ. The green checkmarks indicate the models that visually best reproduce the observations.
\label{fig:CO_abundance}}
\end{figure*}

  We do not attempt to precisely match the brightness of the emitting upper and lower surfaces or the midplane. However, we find that models with a depletion factor between $10^{-3}$ and $10^{-2}$ (\emph{i.e.}, an abundance of $10^{-7}$--$10^{-6}$ relative to H where the temperature is below 20\,K) reproduce the emission significantly better than models assuming complete freeze-out or no freeze-out (Fig.~\ref{fig:CO_abundance}). We note that this range is only indicative, as we varied the freeze-out fraction for a single hydrodynamical model and did not explore potential correlations with, for instance, the disk density and thermal structures, the overall CO abundance, or the amount of UV radiation. Moreover, we did not aim to achieve a perfect match to the data. More quantitative constraints would require a dedicated fit to the data, as done, for instance, by \cite{Hardiman_exoALMA} for DM~Tau. Interestingly, the range we obtain is comparable to the depletion factor of $\approx 1/300$ derived by \cite{Qi2024} for HD~163296 by modeling the optically thin C$^{17}$O and C$^{18}$O $J=1-0$ and $J=2-1$ lines.

\section{Discussion}

The deep exoALMA observations provide an unprecedented view of the $^{12}$CO distribution in protoplanetary disks.  The channel maps show that CO only partially freezes out in the midplane, even when the temperature is below 20\,K, and our simple modeling suggests a relative abundance of the order of $10^{-3}$--$10^{-2}$ compared to the CO abundance in the emitting layers. The midplane abundance we derive is a bit higher than the value found by \cite{Pinte2018a} in IM Lupi (but that study was at a lower spatial resolution of $\approx 0\farcs4$). Desorption at large radii is also systematically required to reproduce the channel maps. The inferred abundance from the observations is generally consistent with thermo-chemical models \citep[e.g.][]{Woitke2009,Bruderer2012}, and more detailed comparisons with models offer a path to better understand photo-desorption processes and the role of UV radiation in shaping the outer disk.

As highlighted by \cite{Dullemond2020}, $^{12}$CO emission from the midplane could potentially be used to directly measure the disk midplane temperature. This method was initially developed for the disk around the Herbig Ae star HD~163296, where most of the disk midplane is above 25\,K, and thus CO is not frozen out, allowing the line to reach optical depths greater than unity. This is likely not the case for the T Tauri stars in our sample. Nevertheless, the systematic detection of midplane CO emission may offer a way to extend the mapping of the disk thermal structure presented by \cite{Galloway_exoALMA} to lower altitudes, particularly in combination with $^{13}$CO to better constrain the optical depths.

Several sources in the exoALMA sample show evidence of embedded planets. Comparisons with models suggest that this is likely the case for LkCa~15, J1615, and J1842, with planet masses estimated to range from one to five Jupiter masses. There is also potential kinematic evidence for planets in SY~Cha and AA~Tau, but reaching a definitive conclusion is more challenging, due either to simplifications in our model for SY~Cha, which does not accurately capture the shape of the velocity signature, or to a limited signal-to-noise ratio for AA~Tau.
We note that while the single-planet model can reproduce the observed velocity kinks, the models are not perfect and do not account for all the detected kinematic structures. This is the case, for instance, in LkCa~15, where additional deviations appear across multiple channels.
  This is illustrated in Fig.~\ref{fig:cont}, where we overlay the expected location of the planet wake \citep{Rafikov2002}.  We use here the same thermal structure as in the {\sf phantom} model, and project the wake vertically from the midplane onto the emitting surface detected with {\sf dynamite}). While some substructures align with the wake's location, additional substructures are observed in the data.

One major limitation of our models is the assumption of a vertically isothermal profile in {\sf phantom}. Comparisons between live radiation and vertically isothermal models showed only minor differences in the kinematic signatures for the Herbig~Ae star HD~97048 \citep{Pinte2009}, but surface thermal waves \citep[e.g.,][]{Dalessio99} or buoyancy spirals \citep{Bae2021} are more likely to develop around lower-mass T~Tauri stars.
Other physical processes can also produce deviations from Keplerian rotation, including magneto-rotational instability (MRI), GI, and vertical shear instability \citep[VSI; see, for instance,][for a comparison]{Barraza_exoALMA}.  The filamentary structures we see in some sources resemble some predictions of VSI or MRI. None of the exoALMA sources seem to display kinematic signatures of GI \citep{Hall2020,Longarini2021}.   We also note that the interplay between instabilities and planets can also significantly affect the expected kinematic signatures, as shown by \cite{Rowther2020} in the case of GI.
A more systematic comparison with models would be necessary to determine whether alternative explanations or additional physics (e.g., vertical thermal structures, additional planets, or instabilities) are required to fully match the observations.

We find planet orbital separations between 80 and 310\,au, comparable to the 260\,au of HD~163296b also detected via disk kinematics \citep{Pinte2018b}. The exoALMA sample is biased toward large and bright disks, but our results (six sources with evidence for planets in a sample of 15 sources) indicate that the presence of planets at large separation is not rare if the disk is large enough. By deprojecting the velocity kinks to the midplane (Fig.~\ref{fig:cont}), we find that most of the potential planets are located just outside the dust continuum emission \citep{Curone_exoALMA}, indicating that they may be truncating the dusty disk. For AA~Tau, the deprojected kink location falls in the dust continuum gap at 80\,au, similar to HD~97048b \citep{Pinte2009}, and the planet candidates in the DSHARP sample \citep{Pinte2020}. The planet candidates are also outside the detected scattered light  \citep{deBoer2016,Ren2023,Ginski2024} and do not seem to be directly related to the structures seen in the near-infrared images, except maybe for J1615 where the candidate is located just outside the outer ring.

The complexity of the observed structures, particularly the filaments between surfaces shown in Fig.\ref{fig:filaments}, suggests that accurately constraining fine parameters, such as the degree and origin of non-thermal broadening, will likely require a precise representation of the underlying emission from the disk. This is especially intriguing in the case of DM~Tau, where non-zero turbulent broadening was inferred from intermediate-resolution ($0\farcs4$) CO observations \citep{Flaherty2020}. The exoALMA data for DM~Tau shows large-scale diffuse emission, striated structures, and non-zero CO emission from the midplane, which may result in line broadening at lower resolution due to smearing. We refer to \cite{Hardiman_exoALMA} for a detailed discussion of non-turbulent broadening in DM~Tau.

Despite the improved data quality provided by exoALMA, the low-level kinematic deviations seen in most of the disks -- and the potential additional physical processes associated with them --
currently set the lower limit for the smallest planet masses we can detect through disk kinematics at $\approx$1\,M$_\mathrm{Jup}$. Deviations caused by lighter planets are of the same order or smaller than this background of deviations, and pushing down our planet detection limit will require more detailed models, for instance accounting for vertical temperature and velocity gradients, self-gravity \citep[e.g.][]{Longarini_exoALMA} or collisional broadening \citep[e.g.][]{Yoshida_exoALMA}.

\section{Conclusions}

We present an in-depth analysis of $^{12}$CO channel maps across 15 protoplanetary disks from the exoALMA survey. Our study reveals a variety of complex kinematic structures, highlighting
the intricate interplay between gas dynamics, including potential planetary interactions, and disk physics and chemistry. The key findings of this work are summarized as follows.

\begin{enumerate}

\item We detect velocity deviations from Keplerian rotation in 13 out of the 15 disks, which manifest in a range of kinematic structures, including extended arcs, spirals, velocity kinks, and filamentary structures.

\item Kinematic signatures observed in six disks point toward the presence of embedded planets.  Our preliminary hydrodynamic and radiative transfer simulations suggest that the planets responsible for the observed velocity kinks have masses in the range of 1 to 5 M$_{\rm Jup}$, and orbital distances between 80 and 310\,au.

\item  Additional physics (\emph{e.g.} vertical temperature and velocity gradients, multiple planets, or disk instabilities) may be required to fully explain the observed deviations from Keplerian motion. Our models are limited in their ability to place tight constraints on planet masses due to these additional complexities. The background of deviations in the observed velocity fields may hide low-mass embedded planets and currently sets the lower limit for the kinematic detection of planets to $\approx$ 1\,M$_\mathrm{Jup}$.

\item The complexity and amplitude of some of the observed kinematic structures, particularly in systems such as MWC~758, suggest that more massive bodies or interactions with the surrounding environment (e.g., infall or outflows) are contributing to the observed deviations.

\item The vertical CO snowline is clearly detected in seven disks where the upper and lower emission surfaces are well separated. Our results indicate partial $^{12}$CO freeze-out in the midplane, with a depletion factor of $\approx 10^{-3}$--$10^{-2}$ compared to the warm molecular layer. Additionally, we systematically detect desorption signatures in the outer regions of these disks.

\end{enumerate}

\bibliography{exoALMA,biblio}
\bibliographystyle{aasjournal}

\section*{Acknowledgments}

This paper makes use of the following ALMA data: ADS/JAO.ALMA\#2021.1.01123.L. ALMA is a partnership of ESO (representing its member states), NSF (USA) and NINS (Japan), together with NRC (Canada), MOST and ASIAA (Taiwan), and KASI (Republic of Korea), in cooperation with the Republic of Chile. The Joint ALMA Observatory is operated by ESO, AUI/NRAO and NAOJ.
The National Radio Astronomy Observatory and Green Bank Observatory are facilities of the U.S. National Science Foundation operated under cooperative agreement by Associated Universities, Inc.
We thank the North American ALMA Science Center (NAASC) for their generous support including providing computing facilities and financial support for student attendance at workshops and publications.
This work was supported by resources awarded under Astronomy Australia Ltd's ASTAC merit allocation scheme on the OzSTAR national facility at Swinburne University of Technology. The OzSTAR program receives funding in part from the Astronomy National Collaborative Research Infrastructure Strategy (NCRIS) allocation provided by the Australian Government, and from the Victorian Higher Education State Investment Fund (VHESIF) provided by the Victorian Government.
C.P. acknowledges funding from the Australian Research Council via FT170100040, DP180104235, and DP220103767.
JB acknowledges support from NASA XRP grant No. 80NSSC23K1312. MB, DF, JS have received funding from the European Research Council (ERC) under the European Union’s Horizon 2020 research and innovation programme (PROTOPLANETS, grant agreement No. 101002188). Computations by JS have been performed on the `Mesocentre SIGAMM' machine, hosted by Observatoire de la Cote d’Azur. PC acknowledges support by the Italian Ministero dell'Istruzione, Universit\`a e Ricerca through the grant Progetti Premiali 2012 – iALMA (CUP C52I13000140001) and by the ANID BASAL project FB210003. SF is funded by the European Union (ERC, UNVEIL, 101076613), and acknowledges financial contribution from PRIN-MUR 2022YP5ACE. MF is supported by a Grant-in-Aid from the Japan Society for the Promotion of Science (KAKENHI: No. JP22H01274). JDI acknowledges support from an STFC Ernest Rutherford Fellowship (ST/W004119/1) and a University Academic Fellowship from the University of Leeds. Support for AFI was provided by NASA through the NASA Hubble Fellowship grant No. HST-HF2-51532.001-A awarded by the Space Telescope Science Institute, which is operated by the Association of Universities for Research in Astronomy, Inc., for NASA, under contract NAS5-26555. CL has received funding from the European Union's Horizon 2020 research and innovation program under the Marie Sklodowska-Curie grant agreement No. 823823 (DUSTBUSTERS) and by the UK Science and Technology research Council (STFC) via the consolidated grant ST/W000997/1. GR acknowledges funding from the Fondazione Cariplo, grant no. 2022-1217, and the European Research Council (ERC) under the European Union’s Horizon Europe Research \& Innovation Programme under grant agreement no. 101039651 (DiscEvol). FMe received funding from the European Research Council (ERC) under the European Union’s Horizon Europe research and innovation program (grant agreement No. 101053020, project Dust2Planets). NC received funding from the European Research Council (ERC) under the European Union Horizon Europe research and innovation program (grant agreement No. 101042275, project Stellar-MADE). H-WY acknowledges support from National Science and Technology Council (NSTC) in Taiwan through grant NSTC 113-2112-M-001-035- and from the Academia Sinica Career Development Award (AS-CDA-111-M03). GWF acknowledges support from the European Research Council (ERC) under the European Union Horizon 2020 research and innovation program (Grant agreement no. 815559 (MHDiscs)). GWF was granted access to the HPC resources of IDRIS under the allocation A0120402231 made by GENCI. Support for BZ was provided by The Brinson Foundation. Views and opinions expressed by ERC-funded scientists are however those of the author(s) only and do not necessarily reflect those of the European Union or the European Research Council. Neither the European Union nor the granting authority can be held responsible for them.

\software{casa \citep{casa}, casa\_cube \citep{casa_cube}, dynamite \citep{Pinte2018a}, phantom \citep{Price2018}, mcfost \citep{Pinte2006,Pinte2009}, pymcfost \citep{pymcfost}}

\end{document}